\documentclass[aps,showkeys,showpacs]{revtex4}
\usepackage{dcolumn}

\def\dddot#1{\mathinner{\buildrel\vbox{\kern5pt\hbox{...}}\over{#1}}}
\def\n{\nonumber}

\def\e{{\rm e}}

\def\d{{\rm d}}

\def\be{\begin{equation}}
\def\ee{\end{equation}}
\def\bq{\begin{eqnarray}}
\def\eq{\end{eqnarray}}
\def\beq{\begin{eqnarray*}}
\def\eeq{\end{eqnarray*}}
\def\p{\partial}

\begin{document}
\title{Gravitational Collapse of Null Radiation and a String fluid}
\author{K S Govinder}\thanks{govinder@nu.ac.za}
\author{M Govender}\thanks{govenderm43@nu.ac.za}
\affiliation{School of Mathematical and Statistical Sciences, University of Natal, Durban 4041, South Africa}
\date{\today}

\begin{abstract}
We consider the end state of collapsing null radiation with a string fluid.  It is shown that, if diffusive transport is assumed for the string, that a naked singularity can form (at least locally). The model has the advantage of not being asymptotically flat.  We also analyse the case of a radiation-string two-fluid and show that a locally naked singularity can result in the collapse of such matter.  We contrast this model with that of strange quark matter.
\end{abstract}



\maketitle
\section{Introduction}
The end state of a collapsing star has attracted much attention in its own right and especially so since the statement of 
the Cosmic Censorship Conjecture (CCC) (See \cite{joshi} for a detailed discussion of the CCC.).  While the end state of a sufficiently massive collapsing star is a singularity \cite{he}, there is no guarantee that such a singularity should be hidden as is required by the CCC \cite{penrose}.  Indeed, a number of papers have been written describing various scenarios in which naked singularities form (See \cite{joshi2} for a recent review.).  Such examples clearly violate the CCC.  Future formulations of the CCC will have to be carefully re-considered in light of these examples.

Here we present another spherically symmetric spacetime which admits a naked singularity.  This spacetime is not asymptotically flat and describes the collapse of null radiation with a string fluid first introduced by Glass and Krisch \cite{gk}.  A full physical description of the model is contained in \cite{gk1}.  We merely quote the relevant information for our analysis of the singularities.

However, we do make one observation:  In their analysis Glass and Krisch \cite{gk,gk1} state that the matter content of their model is a radiation-string two fluid.  The matter content is taken to be a string fluid together with Vaidya null (pure) radiation. We proceed in a similar manner for the main part of our paper.  However, in \S \ref{rflat} we consider the special case of Ricci flat spacetimes.  Here, we do not assume that the string satisfies a diffusion equation. The combination of these two characteristics (Ricci flatness and non-diffusing strings) now allows the matter content to be treated as a different radiation-string two fluid (or, perhaps, a three fluid), where, in addition to the Vaidya null (pure) radiation, we also have incoherent radiation (as the radiation fluid equation of state $p=\rho/3$ is now satisfied) together with the string fluid. 

This note is organised as follows:  In \S \ref{gksoln} we introduce the metric of Glass and Krisch \cite{gk} and present the main equations to solve.  We make a few comments on these equations and provide some special solutions.  These solutions are used in \S \ref{nature} to study the nature of the collapse.  We show that one solution results in a black hole while the other could result in a naked singularity. The strength of the naked singularity is examined in \S \ref{strength}.  During the analysis, we present the general forms of the Kretschmann and Ricci scalars.  We investigate the collapse of Ricci flat spacetimes in \S \ref{rflat} and show that the presence of a string fluid can lead to the occurrence of locally naked singularities for radiation fluids. We compare our results to a similar metric that has recently appeared in the literature in \S \ref{disc}.

\section{The Null Radiation--String Fluid Model} \label{gksoln}
Glass and Krisch \cite{gk}  start with a spherically symmetric metric in  retarded time co-ordinate, $u$, of the form
\be \d s^2 = A \d u^2 + 2 \d u \d r - r^2 (\d \vartheta^2 +\sin^2 \vartheta \d \varphi^2), \label{metric} \ee
where 
\be A = 1 - \frac{2 m(u,r)}{r}. \ee
Note that they chose the mass function to be a function of both retarded time and distance along the outgoing null geodesics. As a result, they were able to extend the Vaidya \cite{vaidya} radiating metric to include a string fluid. 

{\it Remark:} A spherically symmetric line element in advanced time $u$ is usually taken to be of the form 
\be \d s^2 = -A \d u^2 + 2 \d u \d r + r^2 (\d \vartheta^2 +\sin^2 \vartheta \d \varphi^2), \label{metric1} \ee
Thus, the difference between (\ref{metric}) and (\ref{metric1}) is not only a signature change, but also a change of sign of $u$.  We will comment further on this later.

Glass and Krisch \cite{gk}  introduced the unit vectors $\hat v^\mu, \hat r^\mu, \hat\vartheta^\mu$ and $\hat\varphi^\mu$ such that
\be g_{\mu\nu} = \hat v_\mu \hat v_\nu - \hat r_\mu \hat r_\nu - \hat\vartheta_\mu \hat\vartheta_\nu - \hat\varphi_\mu\hat\varphi_\nu,\ee
where the unit vectors have their usual definition from (\ref{metric}).  The Einstein tensor then took on the form
\be G_{\mu\nu} = \frac{2\dot{m}}{r^2}l_\mu l_\nu - \frac{2m'}{r^2}(\hat v_\mu \hat v_\nu - \hat r_\mu \hat r_\nu) + \frac{m''}{r}(\hat\vartheta_\mu \hat\vartheta_\nu + \hat\varphi_\mu\hat\varphi_\nu), \ee
where $'$ and ${}^\cdot$ denote differentiation with respect to $r$ and $u$ respectively and $l_\mu$ is a principal null geodesic vector, from which the nature of the matter content is easily seen.

Using the Einstein field equations they identified the matter portion of the Einstein tensor as a string fluid:
\be T_{\mu\nu} = \psi l_\mu l_\nu + \rho \hat{v}_\mu \hat{v}_\nu + p_r \hat{r}_\mu\hat{r}_\nu + p_\perp(\hat\vartheta_\mu \hat{\vartheta}_\nu+\hat\varphi_\mu\hat\varphi_\nu) \label{emtens} \ee and thus obtained
\bq
4 \pi \psi &=& - \frac{\dot{m}}{r^2} \label{efe1}\\
4 \pi \rho &=& - 4 \pi p_r = \frac{m'}{r^2} \label{efe2} \\
8 \pi p_\perp &=& - \frac{m''}{r}, \label{efe3}
\eq
where $\psi$ is the energy density of the Vaidya radiation, $\rho$ is the string fluid energy density and $p_r$ and $p_\perp$ the string fluid thermodynamic radial and tangential pressures respectively.

In addition to the field equations (\ref{efe1})--(\ref{efe3}) they also assumed that the density $\rho$ satisfied the diffusion equation
\be \dot{\rho} = \frac{\cal D}{r^2} \frac{\p\ }{\p r} \left(r^2\frac{\p \rho}{\p r}\right), \label{diffeqn} \ee
where ${\cal D}$ is the positive coefficient of self-diffusion.  Manipulating equations (\ref{efe1})--(\ref{diffeqn}) they concluded that, once a solution to the linear {\sc pde} (\ref{diffeqn}) was obtained, the mass function could be found by integrating (\ref{efe2}) and
\be \dot{m} = 4 \pi {\cal D} r^2 \frac{\p \rho}{\p r}. \label{newmeqn} \ee
They presented a few solutions of (\ref{diffeqn}) and determined the corresponding expressions for $m$.  

Equation (\ref{newmeqn}) is a result of the integrability conditions taking (\ref{efe1}), (\ref{efe2}) and (\ref{diffeqn}) into account.  However, it would seem that an additive function of $u$ was omitted in (\ref{newmeqn}) which should be 
written as (See Appendix \ref{appa}.)
\be \dot{m} = 4 \pi {\cal D} r^2 \frac{\p \rho}{\p r} + k(u). \label{corrnewmeqn} \ee
However, in our calculations, we will still use (\ref{newmeqn}) and merely add on an arbitrary $k(u)$.

We observe that (\ref{diffeqn}) can be transformed to the heat equation with a nonconstant diffusion coefficient if we set $R=-1/r$.  Thus all the known solutions to the heat equation can be applied in this setting.

It is easy to see that, when $\rho=R(r)U(u)$ we obtain
\be \rho = \e^{u\,\lambda }\,\left( \frac{C_1}
     {\e^{r\,{\sqrt{\frac{\lambda }{\cal D}}}}\,r} + 
    \frac{C_2\e^{r\,{\sqrt{\frac{\lambda }{\cal D}}}}
       }{2\,r\,
       {\sqrt{\frac{\lambda }{\cal D}}}} \right), \label{sol1}\ee
where $C_1$ and $C_2$ are constants and $\lambda$ is the separation constant.  This leads to the following mass function:
\bq m(u,r) &=& \frac{-4\,{\cal D}\,\e^
      {u\,\lambda  - r\,{\sqrt{\frac{\lambda }{{\cal D}}}}}\,
     \pi \,{C}_1}{\lambda } - 
  \frac{4\,{\cal D}\,\e^
      {u\,\lambda  - r\,{\sqrt{\frac{\lambda }{{\cal D}}}}}\,
     \pi \,r\,{\sqrt{\frac{\lambda }{{\cal D}}}}\,
     C_1}{\lambda } \n \\
&&\mbox{}+ 
  \frac{2\,{\cal D}\,\e^
      {u\,\lambda  + r\,{\sqrt{\frac{\lambda }{{\cal D}}}}}\,
     \pi \,r\,C_2}{\lambda } - 
  \frac{2\,{\cal D}^2\,\e^
      {u\,\lambda  + r\,{\sqrt{\frac{\lambda }{{\cal D}}}}}\,
     \pi \,{\sqrt{\frac{\lambda }{{\cal D}}}}\,C_2}
     {{\lambda }^2}. \label{msol1} \eq

If we require that $\rho=R(r)+ U(u)$ then
\be \rho = C_0+\lambda u + \frac{C_1}{r} + \frac{\lambda r^2}{6 \cal D} \label{sol2} \ee
with corresponding mass function
\be m(u,r)=\frac{2\,\pi \,r^5\,\lambda }{15\,\cal D} - 
  4\,{\cal D}\,\pi \,u\,C_1 + 
  \frac{2\,\pi \,r^2\,( 2\,r(u\,\lambda  + C_0)+3C_1)}{3}
       +C_2. \label{msol2} \ee
If we let
\be M(u)=- 4 {\cal D} \pi C_1 u \label{vmass}\ee
be the Vaidya mass and note that the $4\pi C_0r^3/3$ term makes a contribution similar to the cosmological constant, we can define
\be S(u,r) =  \frac{2\,\pi \,r^5\,\lambda }{15\,\cal D}  + 
  \frac{2\,\pi \,r^2\,( 2\,\lambda r u+3C_1)}{3}
       +C_2 \ee
as a mass contribution from the string fluid.  We note that both $S(u)$ and $M(u)$ have a term involving $C_1$.  Thus, in this model, it is not possible to separate the Vaidya mass from the string mass.  This is a result of the relationships (\ref{efe2}) and (\ref{newmeqn}) which are a consequence of (\ref{diffeqn}).

We note that, if we require the string to behave like a perfect fluid, then we must have
$p_r=p_\perp =p$ and so $p=-\rho$ remains as the equation of state for the string fluid. Using equations (\ref{efe2}) and (\ref{efe3}), this means that
\be r m'' - 2 m' = 0 \ee
whence
\be m = F(u) r^3 + G(u). \ee
If we further impose (\ref{diffeqn}), we have that $F(u)=F_0$, a constant and the resulting part of the mass function, ie. $F_0 r^3$ is a cosmological constant term -- the metric is Vaidya-de Sitter (See \cite{wm} for a discussion of the CCC in this context.).  In the resulting discussion, we will assume that the fluid is anisotropic.

\section{Nature of the collapse} \label{nature}
The geodesic equations (taking the null condition $K^aK_a$ into account, where $K^a=\d x^a/\d k$ is the tangent vector of a geodesic) for the metric (\ref{metric}) are
\bq \frac{\d^2u}{\d k^2} + \frac{m-m'r}{r^2}\left(\frac{\d u}{\d k}\right)^2 &=& 0\label{ge1} \\
\frac{\d^2r}{\d k^2} - \frac{\dot{m}}{r}\left(\frac{\d u}{\d k}\right)^2 &=& 0.\label{ge2}
\eq
To investigate the structure of the collapse we need to consider  the radial null geodesics defined by $\d s^2 =0$ taking $\dot\phi=\dot\theta=0$ into account.  For (\ref{metric}) the radial null geodesics must satisfy the null condition
\be \frac{\d u}{\d r} = -\frac{2}{1-\frac{2 m(u,r)}{r}}. \label{rng} \ee
However, we recall that the co-ordinate $u$ is a {\em retarded} time co-ordinate.  As remarked earlier, in most examples in the literature, an {\it advanced} time co-ordinate is used.  In order to compare our results with those usually found in the literature we need to set $u\rightarrow-u$ in (\ref{rng}) and obtain
\be  \frac{\d u}{\d r} = \frac{2}{1-\frac{2 m(-u,r)}{r}}. \label{rng1} \ee
For appropriate forms of $m$, this equation will have a singularity at $r\rightarrow0,u\rightarrow0$.  In order to classify the radial and non-radial outgoing non-spacelike geodesics terminating at this singularity in the past, we need to consider the limiting values of $X=u/r$ along a singular geodesic as the singularity is approached \cite{joshi}.  Thus, for the geodetic tangent to exist uniquely at this point we must have that
\be X_0 = \lim_{u\rightarrow0\ r\rightarrow0}\frac{u}{r} = \lim_{u\rightarrow0\ r\rightarrow0} \frac{\d u}{\d r} =
\lim_{u\rightarrow0\ r\rightarrow0} \frac{2}{1-\frac{2 m(-u,r)}{r}}. \label{gte} \ee
Thus the crucial aspect of metric (\ref{metric}) that dictates the nature of the collapse is, as expected, the mass function $m$.

We note that $\rho=\rho_0$, a constant, is a solution of (\ref{diffeqn}).  The corresponding mass function is $m(u,r)= m_0 + 4 \pi \rho_0 r^3/3$.  For this form of $m$, we have
\be X_0 = \lim_{u\rightarrow0\ r\rightarrow0} \frac{2}{1- \frac{2 m_0 + 8 \pi \rho_0 r^3/3}{r}} = \frac{2}{1-\lim_{r\rightarrow0}\frac{2m_0}{r}}=0. \label{gte1a}\ee
Since there is no real, positive $X_0$, there is no non-spacelike geodesic emanating from the singularity and so the singularity is not visible to any observer.  However, this is not a realistic model as $m$ is independent of time and so the model is static.  The reason we comment on this form of $m$ is to show that any constant part of $m$ will result in a black hole.  As a result in our further analysis of the structure of the collapse we will always set $m_0=0$.

Let us consider solution (\ref{msol1}).  It is clear from that solution that, as $r,u\rightarrow0$, we have $m\rightarrow k$, a constant and so, from (\ref{gte1a}) that a black hole forms.  

In the case of (\ref{msol2}) (with $C_2=0$)
 we observe that (\ref{gte}) reduces to  
\be X_0 = \frac{2}{1-8\pi{\cal D}C_1 X_0} \ee
from which we obtain
\be X_0 = a_\pm=\frac{1\pm\sqrt{1-64 C_1 \pi {\cal D}}}{16 C_1 \pi {\cal D}}. \label{nullcond} \ee
If we define $\delta= 8C_1\pi{\cal D}$ we can rewrite (\ref{nullcond}) as
\be X_0 = a_\pm=\frac{1\pm\sqrt{1-8\delta}}{2\delta}.\label{nullcond1} \ee
In order for real positive solutions in (\ref{nullcond1}) to exist we must have that  
\be \delta \leq \frac18.  \label{deltacond}\ee
 This is exactly the same form of $X_0$ (and restriction) obtained by Dwivedi and Joshi \cite{dj} and Wagh and Maharaj \cite{wm}.  That the similarity arises should not be of any surprise.  It is clear that, if one is dealing with a mass function that is of the form $\lambda u+F(u,r)$ where $F(u,r)/r$ is polynomial in $r$ and/or $u$ (of degree at least one), then the only contribution to the condition for the existence of a time-like geodesic emanating from the singularity is from $\lambda u$, the Vaidya mass.  Indeed, this is the reason that asymptotic flatness is not essential for the existence of a locally naked singularity, a conclusion first obtained in \cite{wm}.

\section{Strength of the singularity} \label{strength}
The main importance of determining the strength of the singularity is due to the fact that the CCC does not need to rule out the possibility of the occurrence of weak naked singularities \cite{nolan}.  This arises as one may continue the geometry of a weak naked singularity through the singularity to make it geodesically complete \cite{es,ori} (See also \cite{nolan}.).

In order to determine the strength of the singularity we utilise the ideas of Nolan \cite{nolan} as explained in \cite{gd}.   Thus, for a weak singularity, we require
\be \frac{\d r}{\d k} \sim d_0 \qquad \Rightarrow r \sim d_0 k. \label{sc1} \ee
Using the definition 
\be X_0 = \lim_{u\rightarrow0\ r\rightarrow0}\frac{u}{r} = \lim_{u\rightarrow0\ r\rightarrow0} \frac{\d u}{\d r} \ee
we have, asymptotically,
\be \frac{\d u}{\d k} \sim d_0 X_0 \qquad \Rightarrow u \sim d_0 X_0 k. \label{sc2} \ee
The geodesic equation (\ref{ge1}) then becomes 
\be \frac{d^2u}{dk^2}\sim
\frac{2X_0^2 \pi d_0}{15 k {\cal D}} (4 \lambda d_0^4 k^4+20 d_0^3 k^3 d X_0 \lambda+20 d_0^2 k^2 {\cal D} C_0+15 d_0 k {\cal D} C_1+30 d^2 X_0 C_1) \label{sc3}
\ee
which is of $O(k)^{-1}$ and so is inconsistent with (\ref{sc2}).  Thus the singularity is gravitationally strong in the sense of Tipler \cite{tip}.

The Kretschmann scalar $K=R_{abcd}R^{abcd}$ for (\ref{metric}) is given by
\be K = \frac{4}{r^6}((m'')^2 - 4 m'' m' r^3 + 4 m m'' r^2 + 8 (m')^2 r^2 - 16 m m' r + 12 m^2). \label{kresch1} \ee
If we utilise the mass function (\ref{msol2}), (\ref{kresch1}) evaluates to
\beq K&=&\frac{64\pi^2}{225{\cal D}^2 r^6} (1200 r^6{\cal D}^2 u \lambda C_0+600 r^6{\cal D}^2 u^2 \lambda^2+900 r^5{\cal D}^2 C_0 C_1+300 r^8 \lambda^2{\cal D} u+300 r^8 \lambda{\cal D} C_0+150 r^7 \lambda{\cal D} C_1\n \\
&&\mbox{}+2700{\cal D}^4 u^2 C_1^2+720 r^5 \lambda{\cal D}^2 u C_1+600 r^6{\cal D}^2 C_0^2+450 r^4{\cal D}^2 C_1^2+53 r^{10} \lambda^2)\eeq which clearly diverges at the naked singularity.  Hence the singularity is a scalar polynomial singularity.  

The Ricci scalar for (\ref{metric}) is given by
\be R = \frac{-2}{r^2}(m''r+2m') \label{ricci1}\ee
which, in the case of (\ref{msol2}) evaluates to
\be R=\frac{-8\pi}{{\cal D} r}(r^3\lambda+4{\cal D}ru\lambda+4{\cal D}rC_0+3{\cal D}C_1)\ee
and also diverges at the naked singularity.

\section{Ricci Flat Spacetimes} \label{rflat}

It is clear from (\ref{ricci1}) that none of the solutions presented above cover the case of Ricci flat spacetimes.  If we relax the requirement that the string diffuses (ie. we do not impose (\ref{diffeqn})) we can now consider the case of $R=0$.  From (\ref{ricci1}) this means that the mass function must take on the form
\be m(u,r) = \bar{M}(u)+\frac{\bar{S}(u)}{r}, \label{rflatmass} \ee
where we can again interpret $\bar{M}(u)$ as the Vaidya mass and $\bar{S}(u)$ as the mass contribution from the matter content. Forcing the spacetime to be Ricci flat requires the matter content to obey the (incoherent) radiation fluid equation of state, ie.
\be p\equiv \frac{p_r+2p_\perp}{3}=-\frac{p_r}{3} = \frac{\rho}{3} \label{radeos1} \ee
from (\ref{efe2}) and (\ref{efe3}).  Thus if we begin with a metric of the form (\ref{metric}) (together with an energy--momentum tensor of the form (\ref{emtens})) and require it to be Ricci flat, then, as expected, the fluid 
must satisfy (\ref{radeos1}).  These spacetimes must be asymptotically flat via (\ref{rflatmass}).

To proceed further, we must assume some form for $\bar{M}(u)$ and $\bar{S}(u)$.  Given the interpretation above, it makes sense to assume that $\bar{M}(u)=-\alpha u$.  In order for $\bar{S}(u)$ to contribute to the existence (or otherwise) of a naked singularity, it must be of $O(u)^2$.  Thus we set $\bar{S}(u)=\beta u^2$. 
The null condition (\ref{rng1}) now becomes 
\be  \frac{\d u}{\d r} = \frac{2}{1-\frac{2 \alpha u + 2 \beta u^2/r}{r}}. \label{rng1a} \ee
This equation has a singularity at $r\rightarrow0,u\rightarrow0$.  The condition for a geodetic tangent to exist uniquely at this point reduces to
\be X_0 = \lim_{u\rightarrow0\ r\rightarrow0}\frac{u}{r} = \lim_{u\rightarrow0\ r\rightarrow0} \frac{\d u}{\d r} =
\lim_{u\rightarrow0\ r\rightarrow0} \frac{2}{1-\frac{2  \alpha u + 2 \beta u^2/r}{r}}=\frac{2}{1-2\alpha X_0 -2 \beta X_0^2} \label{gte1} \ee
This cubic equation in $X_0$ has (at least) one real positive root provided that, for $\alpha>0$, $\beta<0$.  
This can be seen from the numerical results \cite{mma} given in Table \ref{tab1}.  Thus, in the case of Ricci flat spacetimes, we also have the possibility of a locally naked singularity.

As above, we investigate the strength of the naked singularity.  Taking (\ref{sc1}) and (\ref{sc2}) into account, (\ref{ge1}) (with (\ref{rflatmass}) and the assumed forms of $\bar M(u)$ and $\bar S(u)$) becomes
\be \frac{\d^2u}{\d k^2} \sim \frac{2 \beta X_0 - \alpha }{d_0 k} \ee
which is inconsistent with (\ref{sc2}) and so this singularity is also gravitationally strong.

For Ricci flat spacetimes, the Kretschmann scalar (\ref{kresch1}) is given by 
\be K =  \frac{16}{r^8}(14\bar S(u)^2+12\bar S(u)\bar M(u)r+3\bar M(u)^2r^2) \ee
which, for the assumed forms of $\bar M(u)$ and $\bar S(u)$, evaluates to 
\be K = \frac{16 u^2}{r^8}(14\beta^2u^2-12\beta \alpha u r+3\alpha^2r^2). \label{kfin} \ee
Clearly (\ref{kfin}) diverges at the naked singularity.  As expected, we have a scalar polynomial singularity. 

\begin{table}
\caption{\label{tab1}Positive numerical solutions to (\ref{gte1}) for specific forms of $\alpha=1.0$ and $\beta$.}
\begin{ruledtabular}
\begin{tabular}{dd}
\beta& X_0\\
 -1& 1.35321 \\
 -0.9 & 1.42537 \\
 -0.8 & 1.51454\\
 -0.7 &1.62861\\
 -0.6&1.78145\\
-0.5&2.0\\
 -0.4 &2.3436\\
-0.3&2.96715\\
 -0.2&4.3797\\
 -0.1&9.1457
\end{tabular}
\end{ruledtabular}
\end{table}

\section{Discussion} \label{disc}
We have presented a scenario for the gravitational collapse of a string fluid together with Vaidya null radiation in which the  end state becomes a naked singularity.  We have also investigated the collapse of Ricci flat spacetimes. Such collapse can also give rise to  naked singularities.  Both scenarios provided naked singularities which were gravitationally strong and, as expected, were scalar polynomial singularities. These are two more in the long list of examples that proliferate the literature and should serve to tighten future formulations of the CCC.

We comment briefly on our results and those of Ghosh and Dadhich \cite{gd}.  The metrics (apart from a signature change and change of sign of time co-ordinate) are similar.  As a result, the Einstein field equations also reduce to (\ref{efe1}--\ref{efe3}) (though their $p$ represents isotropic pressure).  However, in their model, Ghosh and Dadhich \cite{gd} interpret the matter content as strange quark matter.  Thus, while $\psi$ is still the radiation density, their $p$ and $\rho$ refer to strange quark matter thermodynamic pressure and energy density respectively. As a result of that interpretation, they impose an equation of state relating $p$ to $\rho$:
\be p = \frac{1}{n}(\rho-4B), \label{eosgd} \ee
where $B$ is the bag constant.  This differential equation can be treated as an {\sc ode} and the general solution was presented in \cite{gd}. 

Here, following Glass and Krisch \cite{gk}, the matter content has been interpreted as a string fluid which diffuses like point particle diffusion where the number density, $n$, diffuses from higher numbers to lower according to (\ref{diffeqn}) (Note we have used the fact that $\rho=M_s n$, $M_s$ a constant,  to obtain (\ref{diffeqn}).).  The string equation of state is always satisfied as a result of the energy-momentum tensor (\ref{emtens}). As a result, the solutions presented here (and in \cite{gk}) will not coincide with those found in \cite{gd}.

That other interpretations of the metric (\ref{metric}) together with (\ref{emtens}) are possible  is evident from the case of Ricci flat collapse investigated in \S \ref{rflat}. Indeed, this case is of considerable importance as the matter content satisfies {\it both} the string equation of state $p_r = -\rho$ and the (incoherent) radiation fluid equation of state $p = \rho/3$ (see (\ref{radeos1})).  Thus the matter content is a two-fluid, in addition to the Vaidya null (pure) radiation.  We have shown that a naked singularity can arise in this scenario. In \cite{cjjs} it was shown that a naked singularity only forms if the ratio of pressure to density was less than $-1/3$. Here, we have shown that a naked singularity can exist outside this range, ie. for a ratio of $1/3$. We conclude that it is the presence of the string matter that contributes to the occurrence of the locally naked singularity.

As observed in \S \ref{gksoln}, an arbitrary function of $u$, $k(u)$ say, can be added to the mass function $m$.  This additive function of $u$ can be seen as a further contribution from the string mass $S$.  However, unless it is linear in $u$, it would not contribute to the condition (\ref{deltacond}) and even then, would only serve to redefine $\delta$.

We note that the usual manner in which the existence of a naked singularity is presented is to show that the singularity forms before the event horizon forms.  Interestingly, it has been shown that, for some models, an apparent horizon will not form at all \cite{bd,h1,k1} (See also \cite{waghetal}.).

\begin{acknowledgments}
We thank Naresh Dadhich, Roy Maartens and Sunil Maharaj for useful discussions and helpful suggestions. 
KSG thanks the University of Natal for continuing support. Many of the tensor calculations have been performed using {\it GRTensor 1.79} \cite{grtens}. We thank the referees for their constructive criticisms that have helped improve this paper.
\end{acknowledgments}

\appendix
\section{Integrability conditions for $m(u,r)$} \label{appa} 

Here we consider the derivation of (\ref{newmeqn}).  Differentiating (\ref{efe1}) and (\ref{efe2}) with respect to $r$ and $u$ respectively results in
\bq 
\dot{m}' &=& - 4\pi (r^2\psi)' \label{app1}\\
\dot{m}' &=& 4 \pi r^2 \dot{\rho}. \label{app2}
\eq
Thus
\be - (r^2 \psi)' = r^2 \dot{\rho}. \label{app3} \ee
However, (\ref{diffeqn}) yields
\be r^2 \dot{\rho} = {\cal D} \frac{\p\ }{\p r} \left(r^2\frac{\p \rho}{\p r}\right). \label{app4} \ee
Thus, from (\ref{app1}), (\ref{app3}) and (\ref{app4}) we have
\be - (r^2 \psi)' = \frac{\dot{m}'}{4 \pi} = {\cal D} \frac{\p\ }{\p r} \left(r^2\frac{\p \rho}{\p r}\right) \ee
whence
\be \dot{m}' = 4 \pi {\cal D} \frac{\p\ }{\p r} \left(r^2\frac{\p \rho}{\p r}\right). \ee
Glass and Krisch \cite{gk} presumably integrated this equation to obtain (\ref{newmeqn}).  However, since this is a {\sc pde}, the integration should yield (\ref{corrnewmeqn}).

\end{document}